\documentclass[twocolumn,superscriptaddress,longbibliography,aps,preprintnumbers]{revtex4-2}

\usepackage{graphicx,color}
\usepackage{bm,bbm,amsmath,amssymb,braket}
\usepackage{epstopdf}
\usepackage{relsize}
\usepackage{epstopdf}
\usepackage{comment}
\usepackage{soul}
\usepackage{float}
\usepackage{lipsum}
\usepackage[urlcolor=blue,colorlinks=true,citecolor=blue,
	linkcolor=blue,pdfstartview={FitH},bookmarks=false]{hyperref}
\usepackage{MnSymbol}
\usepackage{stmaryrd}
\usepackage{mathtools}

\graphicspath{{Figures/}{./Figures/}{.}{./paper_figures/}}
\setstcolor{red}

\definecolor{purple}{rgb}{0.5, 0., 0.8}

\begin{document}
\title{ Entanglement entropy scaling laws from fluctuations of non-conserved quantities}
\author{Szczepan G\L{}odzik}
\affiliation{Institute of Physics, M.\ Curie-Sk\l{}odowska University, 
20-031 Lublin, Poland}
\author{Ali G.  Moghaddam}
\affiliation{Computational Physics Laboratory, Physics Unit, Faculty of Engineering and Natural Sciences, Tampere University, P.O. Box 692, FI-33014 Tampere, Finland}
\affiliation{Helsinki Institute of Physics P.O. Box 64, FI-00014, Finland}
\affiliation{Department of Physics, Institute for Advanced Studies in Basic Sciences (IASBS), Zanjan 45137-66731, Iran}
\author{Kim P\"oyh\"onen}
\affiliation{Computational Physics Laboratory, Physics Unit, Faculty of Engineering and Natural Sciences, Tampere University, P.O. Box 692, FI-33014 Tampere, Finland}
\affiliation{Helsinki Institute of Physics P.O. Box 64, FI-00014, Finland}
\author{Teemu Ojanen}
\affiliation{Computational Physics Laboratory, Physics Unit, Faculty of Engineering and Natural Sciences, Tampere University, P.O. Box 692, FI-33014 Tampere, Finland}
\affiliation{Helsinki Institute of Physics P.O. Box 64, FI-00014, Finland}

\date{\today}

\begin{abstract}
Entanglement patterns reveal essential information on many-body states and provide a way to classify quantum phases of matter. However, experimental studies of many-body entanglement remain scarce due to their unscalable nature. The present work aims to mitigate this theoretical and experimental divide by introducing reduced fluctuations of observables, consisting of a sum of on-site operators, as a scalable experimental probe of the entanglement entropy. Specifically, we illustrate by Density Matrix Renormalization Group calculations in spin chains that the reduced fluctuations exhibit the same size scaling properties as the entanglement entropy. Generalizing previous observations restricted to special systems with conserved quantities, our work introduces an experimentally feasible protocol to extract entanglement scaling laws.
\end{abstract}

\maketitle

\emph{Introduction}.—The traditional way to classify phases of matter relies on broken symmetry and order parameters. More recently, topology has emerged as another fundamental characterization of many-body states \cite{Wen2017,Kitaev2009,Ryu2016}. In the emerging quantum information era, many-body entanglement has been understood as yet another important organization principle of conventional and simulated quantum matter. Despite being the hallmark property of quantum mechanics \cite{Horodecki}, the importance of many-body entanglement for states of matter has been recognized relatively recently \cite{zeng2019book,vedral2008rmp,plenio2014introduction,laflorencie2016quantum,Kitaev2006,Fidkowski2011,calabrese2007entanglement,eisert2015quantum,osterloh2002scaling,Kitaev2003,hastings2007area,Plenio2005,Wolf2006}. Entanglement properties offer a complementary view to symmetry and topology, and in some cases, even the only classifying principle to distinguish different phases of matter \cite{wen2006,Balents2012identifying,Fisher2018,Skinner2019,Fisher2022review}. For example, some quantum states which are indistinguishable by their symmetry and topology, can be distinguished by their different entanglement entropy scaling laws \cite{Zanardi2005,Swingle2010,Nayak2013,Abanin2019rmp}. Symmetry, topology and entanglement provide the lens through which we currently analyze quantum phases of matter.

While consequences of broken symmetry and topology could be experimentally probed through local order parameters and physical responses, there exists no simple and direct experimental probe for entanglement. In fact, a direct attempt to measure  size-scaling properties of entanglement entropy requires quantum state tomography, which scales exponentially in the system size. The root difficulty of experimental studies lies precisely in the unscalable nature of different entanglement measures \cite{Cramer2010tomography,Eisert2010,Greiner2015measuring,Brydges2019probing}. One way to try to circumvent exponential complexity is to try to establish a connection between the entanglement measure of interest and a conveniently accessible observable.  
In particular, connections between entanglement and correlations have been established for quite some time, but primarily in the form of loose lower bounds on entanglement or mutual information \cite{Wolf2007,Verstraete2004}. However, in the presence of conserved quantities and symmetries, a more persuasive connection between the entanglement entropy and the fluctuations of conserved observables has been observed \cite{Klich2006,Klich2009,LeHur2010,LeHur2012}.
These efforts recently culminated in the counting argument, according to which the fluctuations of an extensive conserved observable exhibit the same scaling properties as entanglement entropy \cite{Poyhonen2022}. This argument has been confirmed for many different systems and states which exhibit area-law, volume-law and critical scaling of entropy.
\begin{figure}
    \centering
    \includegraphics[width=\linewidth]{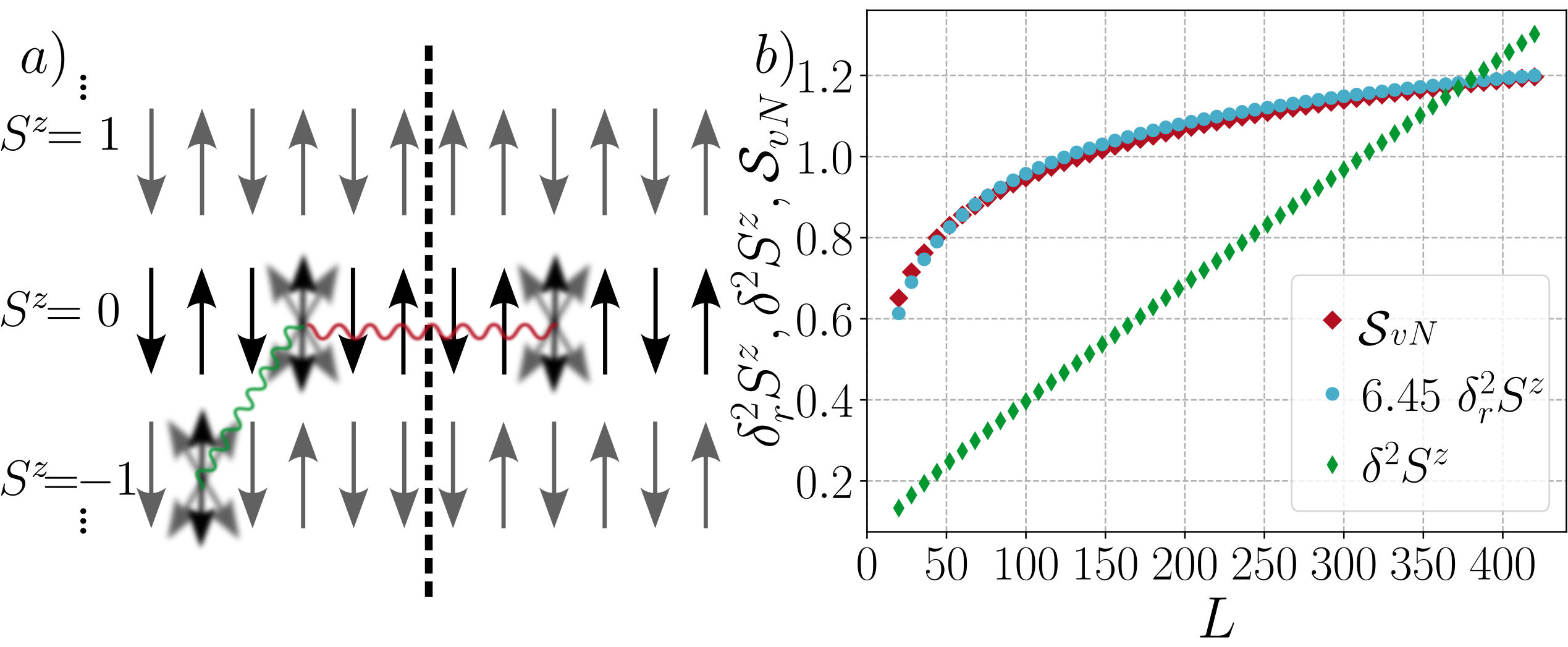}
    \caption{a): Illustration of fluctuation-entanglement relation in spin chains. The subsystem spin fluctuations within a spin sector (red) are sensitive to bipartite entanglement. In contrast, the subsystem fluctuations between sectors (green), taking place when the total spin is not conserved, give rise to a spurious volume-law term independent on entanglement. b): Scaling of subsystem spin fluctuations and entanglement entropy in a critical chain. While the subsystem spin variance exhibits a volume-law scaling due to non-conservation, the reduced fluctuation $\delta_r^2S^z$, introduced in the present work, exhibits the same scaling as the entanglement entropy $\mathcal{S}_{vN}$. 
    The reduced fluctuations are rescaled by a numerical factor for visualization purposes. }
    \label{fig:intro_fig}
\end{figure}

In this paper we address the outstanding problem of how to devise a scalable protocol to extract entanglement entropy scaling laws in the general case when no conserved quantities are available. For non-conserved observables, as illustrated in Fig.~\ref{fig:intro_fig}(a),  the connection between subsystem fluctuations and entanglement is obscured by processes that do not require entanglement. To restore the connection, we introduce a new quantity, the reduced fluctuation of a sum of on-site observables, and argue that it serves as a scalable proxy to extract entanglement entropy scaling laws. As illustrated in Fig,~\ref{fig:intro_fig}(b), Density Matrix Renormalization Group (DMRG) calculations \cite{White1992,Schollwock2005dmrg,verstraete2023} in spin 1/2 chains support our argument that the reduced fluctuations exhibits the same scaling as the entanglement entropy. Our protocol based on reduced fluctuations is particularly suited to studying entanglement scaling laws in quantum simulator systems and Noisy Intermediate-Scale Quantum (NISQ) devices \cite{preskill2018NISQ,Altman2021,google2023measurement,guo2024nature}.

\begin{figure*}
   \centering
    \includegraphics[width=1.05\linewidth]{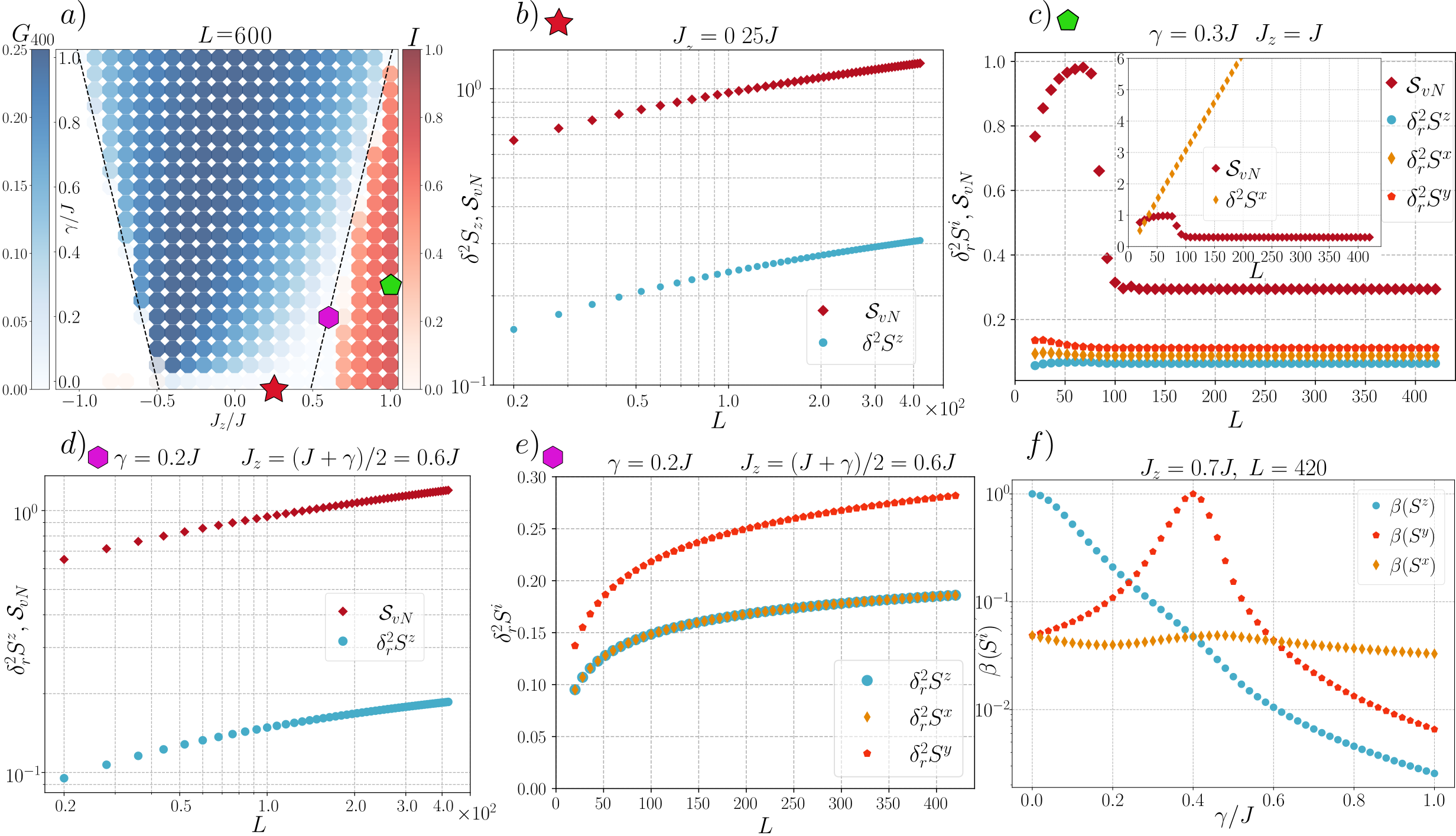}
    \caption{Numerical results: a) Phase diagram in the $\gamma$--$J_z$ space. The critical lines given by $J_z=\pm(J+\gamma)/2$ are presented as dashed lines. Colorful shapes refer to parameters for which other marked figures are obtained. b) Critical (log) scaling of von Neumann entropy and $S^z$ half system fluctuations in the conserved case ($\gamma=0$). c) Area law scaling of von Neumann entropy and the reduced fluctuations of all spin operators. Inset: half system fluctuations of $S^x$ showing volume law scaling due to being non-conserved. d) Critical (log) scaling of von Neumann entropy and scaled reduced $S^z$ fluctuations. e) Same as d) but without entanglement entropy and showing all spin directions. f) Visibility for all spin operators along a cut in the phase diagram $J_z=0.7J$.}
    \label{fig:megafigure}
\end{figure*}

\emph{Reduced fluctuations and entanglement}.—We consider quantum states defined on $D$-dimensional lattice systems. 
 The von Neumann entanglement entropy in state $\ket{\Psi}$ quantifies the entanglement between a subsystem $\Omega$ and its complement $\overline{\Omega}$ and is defined as $\mathcal{S}_{vN}=-\mathrm{Tr}_{\Omega} \rho_\Omega \log \rho_\Omega$, where $\rho_\Omega=\mathrm{Tr}_{\overline{\Omega}}\ket{\Psi}\bra{\Psi}$ is the reduced density matrix in $\Omega$. States arising from local interactions or unitaries exhibit, in the leading order, an area-law $\mathcal{S}_{vN}\propto L^{D-1}$, a volume-law $\mathcal{S}_{vN}\propto L^D $  or a critical scaling $\mathcal{S}_{vN}\propto L^{D-1}\log L$, where $L$ is the characteristic linear dimension of $\Omega$. The purpose of the present work is to introduce a quantity which exhibits the same scaling laws but could be experimentally probed without an exponential complexity associated with $\mathcal{S}_{vN}$. To this end, we consider extensive observables constituting of a sum of on-site terms $\hat{A}_{\rm tot}=\sum_{i}\hat{a}_i$, including subsystem extensive observables $\hat{A}_{\Omega}=\sum_{ i \in \Omega}\hat{a}_i.$ Here $\hat{a}_i$ are on-site operators acting on degrees of freedom at each site $i$. For example, a natural choice for spin systems is an on-site spin operator and for fermionic systems an on-site number operator. 

It has been previously observed \cite{Poyhonen2022} that, for conserved cases where $[\rho_{\Omega}, \hat{A}_{\Omega}] = 0$, the fluctuations of subsystem observables, $\delta^2\hat{A}_{\Omega} = \langle \hat{A}_{\Omega}^2 \rangle - \langle \hat{A}_{\Omega} \rangle^2$,
obey the same scaling laws as the entanglement entropy. This conservation typically arises when the system is in an eigenstate of a Hamiltonian $\hat{H}$ that commutes with the operator $\hat{A}_{\rm tot}$, ($[\hat{A}_{\rm tot}, \hat{H}] = 0$). 
The reason behind the similarity of scaling between entanglement entropy $S_{vN}$ and , for example, spin fluctuations under conservation, can be briefly explained as follows (see \cite{Poyhonen2022} for more details). From the scaling of $S_{vN}$ with subsystem size, one can infer that the number of active degrees of freedom contributing to entanglement scales as $S_{vN} \propto N_a$, where $N_a$ is the number of spins in the subsystem. Under conservation, subsystem spin fluctuations originate purely from the variation in spin configurations, each characterized by a subsystem spin $S_z$. The probability distribution of these configurations can be well approximated by the binomial form $p_{S_z}=\binom{N_a}{S_z} \frac{1}{2^{N_a}}$, leading to a subsystem spin variance also proportional to $N_a$. This shows that both entanglement and spin fluctuations scale with the number of subsystem spins effectively entangled with the rest of the system.

Now, as illustrated in Fig.~\ref{fig:intro_fig}, when $ \hat{A}_{\Omega}$ is not conserved, its fluctuations are no longer linked to quantum correlations between the subsystems. As the number of non-conserved fluctuation processes scale as the subsystem volume, fluctuations $\delta^2\hat{A}_{\Omega}$ exhibit volume-law scaling irrespectively of the entanglement entropy scaling. To extend the applicability of fluctuations as a proxy for the entanglement entropy in generic (non-conserved) systems, we introduce the notion of \emph{reduced fluctuation} as
\begin{align}\label{eq:reduced2}
    \delta_r^2\hat{A} = -\langle \hat{A}_{\Omega} \hat{A}_{\overline{\Omega}} \rangle_c = \langle \hat{A}_{\Omega} \rangle \langle \hat{A}_{\overline{\Omega}} \rangle - \langle \hat{A}_{\Omega} \hat{A}_{\overline{\Omega}} \rangle,
\end{align}
using the connected correlator of the subsystem observable $\hat{A}_{\Omega}$ with its complement observable $\hat{A}_{\overline{\Omega}}$.
\par
To illustrate the utility of the reduced fluctuation as a proxy measure of entanglement, we first note that both the reduced fluctuation and the entanglement entropy vanish for states that are separable, i.e., product states of the form $\ket{\Psi} = \ket{\psi_{\Omega}} \otimes \ket{\psi_{\overline{\Omega}}}$. Consequently, a nonzero value of the reduced fluctuation indicates entanglement between the subsystems \footnote{The converse, however, is not generally true; certain entangled states may also yield zero or even negative values for the reduced fluctuations.}. 
Second, the reduced fluctuation can be expressed as
\begin{equation}\label{eq:reduced4}
    \delta_r^2\hat{A} = \frac{1}{2} \left( \delta^2\hat{A}_{\Omega} + \delta^2\hat{A}_{\overline{\Omega}} - \delta^2\hat{A}_{\rm tot} \right),
\end{equation}
in terms of the variances of the observables for the two subsystems and the total system. The equivalence between the two expressions becomes clear by expanding the variance of the total observable as $\delta^2\hat{A}_{\rm tot} = \langle(\hat{A}_\Omega + \hat{A}_{\bar{\Omega}})^2\rangle - 
\langle\hat{A}_\Omega + \hat{A}_{\bar{\Omega}}\rangle^2$ and inserting it into Eq.~\eqref{eq:reduced4}. All terms except the cross terms between the two subsystems cancel, yielding Eq.~\eqref{eq:reduced2} once again.
The fact that the total system fluctuations—which can be viewed as the volume-law contribution of non-entanglement origin—are subtracted from the subsystem contributions, further justifies the term \emph{reduced fluctuations} (schematically illustrated in Fig.~\ref{fig:intro_fig}(a) for spin chains).
In the special case of symmetric bipartitioning, where $|\Omega| = |\overline{\Omega}|$, the reduced fluctuation further simplifies to 
$\delta_r^2\hat{A} = \delta^2\hat{A}_{\Omega}-(1/2)\delta^2\hat{A}_{\rm tot} $. For the case of conserved quantities, since $[A_{\rm tot},\ket{\Psi}\bra{\Psi}] = 0$, the variance of the total system observable vanishes ($\delta^2\hat{A}_{\rm tot} = 0$). Therefore, the reduced fluctuation simplifies to 
$\delta_r^2\hat{A} = \delta^2\hat{A}_{\Omega}$, which is the same as the variance of the subsystem observable. 
\par
By expanding Eq.~\eqref{eq:reduced2} in terms of on-site operators as 
\begin{equation}\label{eq:reduced3}
 \delta_r^2\hat{A}=-\sum_{i\in \Omega, j\in \overline{\Omega}}\langle\hat{a}_i\hat{a}_j \rangle_c,  
 \end{equation}
it becomes further evident that the reduced fluctuation is sensitive to the entanglement between the subsystems. In particular, for states where the correlations are exponentially suppressed $\langle\hat{a}_i\hat{a}_j \rangle_c\propto e^{-|\vec{r}_i-\vec{r}_j|/\xi}$ with a characteristic correlation length $\xi$ much smaller than the subsystem linear dimensions, only the lattice sites in the subsystem surface layer of width $\sim\xi$ contribute to the sum.  As a consequence, the reduced fluctuation scales as $\delta_r^2\hat{A}\propto \partial \Omega$, where $\partial \Omega$ denotes the surface area between subsystem $\Omega$ and its complement $\overline{\Omega}$. For short-range correlated systems one would also expect area-law scaling for the entanglement entropy, supporting the idea that the reduced fluctuation and the entanglement entropy exhibit the same scaling laws. Below we illustrate in spin chains that this entanglement-fluctuation correspondence holds beyond the area-law case.  
\par
Unsurprisingly, the entropy-fluctuation correspondence works better for some operators and states than others. The sensitivity of the reduced fluctuation to entanglement is determined by the value of the total system variance $\delta^2\hat{A}_{\rm tot}$, which can be regarded as a background noise in observing entanglement. This motivates the definition of the visibility 
\begin{equation}
    \beta=1-\delta^2\hat{A}_{\rm tot}/(\delta^2\hat{A}_\Omega+\delta^2\hat{A}_{\overline\Omega}),
\end{equation} as a measure of sensitivity of the chosen observable to entanglement in the studied state. For conserved quantities, the background noise vanishes and the visibility saturates its maximum $\beta=1$, while for non-conserved quantities the visibility generally satisfies $-1\leq\beta<1$. The visibility limits the reduced fluctuation signal as $\delta_r^2\hat{A}= \beta \delta^2A_\Omega$. For states and operators for which the visibility vanishes, or becomes very small, the reduced fluctuation becomes insensitive to entanglement. The larger the visibility, the more sensitive probe the operator $\hat{A}$ is to entanglement. 

\emph{Spin $1/2$ chains}.—We now illustrate the fluctuation-entanglement correspondence in spin 1/2 XYZ chain
\begin{equation}
    H=-\mathlarger{\sum}\limits_{j}^{L-1}\left[\dfrac{J+\gamma}{2}S^x_jS^x_{j+1}+\dfrac{J-\gamma}{2}S^y_{j}S^y_{j+1}+J_z
    S^z_{j}S^z_{j+1} \right]
    \label{eq:xyz_ham}
\end{equation}
where $J,J_z$ and $\gamma$ parametrize the nearest-neighbor exchange couplings. We use the ITensor package~\cite{itens1,itens2} to find the ground state of the Hamiltonian Eq.~\eqref{eq:xyz_ham} using the DMRG algorithm. At most $50$ sweeps and a maximum bond dimension $\chi=512$ were used, which ensured that the energy variances drop below $10^{-12}$. The phase diagram as a function of $\gamma$ and $J_z$ is illustrated in Fig.~\ref{fig:megafigure}(a). We identify the known three phases of the system marked with colors: (i) the antiferromagnetic phase (red octagons) signalled by the onset of the spin imbalance $I=(\sum_{j_e}\langle S^z_j\rangle - \sum_{j_o}\langle S^z_j\rangle)/(\sum_j\langle S^z_j\rangle + L)$, where $j_{e,o}$ refers to even and odd sites respectively; (ii) the Ising antiferromagnetic (topological superconducting in the fermion language) phase, marked with blue octagons in the phase diagram, and probed by the non-zero value of $G_r=\sum_{j=1}^{L-r}|\langle S^x_j S^x_{j+r}\rangle|/(L-r)$; (iii) the band insulator phase (white region) with both of the aforementioned probes being zero. All three of the recognized phases are gapped, with degenerate ground states. We conserve the spin parity in our calculations, and the presented results are obtained for the even sector. When $\gamma=0$, the system is in the XXZ critical limit, and is symmetric with respect to rotations around $z$ axis, with $S^z=\sum_i S_i^z$ conserved. As observed previously and illustrated in Fig.~\ref{fig:megafigure}(b), in the conserved case $S^z$ fluctuations closely follow the logarithmic scaling of the entanglement entropy.   

For finite anisotropy $\gamma\neq 0$, the system generally supports no conserved spin. In this case, as illustrated for the antiferromagnetic phase in the inset of Fig.~\ref{fig:megafigure}(c), the spin fluctuations exhibit a volume-law scaling while the entanglement entropy obeys the 1d area law. Thus, the behavior of simple fluctuations and the entanglement entropy is decoupled. However, by considering the reduced spin fluctuations $\delta^2_r S^i=\delta^2 S^i(L/2)-\frac{1}{2}\delta^2 S^i(L)$ for $i=x,y,z$, the correspondence between fluctuations and the entanglement entropy is recovered. The area-law scaling in 1D implies that the entanglement entropy becomes independent of the subsystem size as it becomes larger than the correlation length, which determines the decay length of the correlations. As seen in Fig.~\ref{fig:megafigure}(c), the reduced fluctuations settle to a constant value at the same length scale as the entanglement entropy. The difference in low-length behavior of the entanglement entropy is due to a known feature of DMRG and does not occur e.g. in the topological phase where the even spin sector is not degenerate. Thus, in agreement with the argument in the previous section, the reduced fluctuations in the entanglement entropy area-law regime indeed follow area law too.

At the phase boundary between the gapped phases, the system becomes critical. As seen in Fig.~\ref{fig:megafigure}(a), the phase boundaries are determined by the lines $J_z=\pm(J+\gamma)/2$. Each point of the boundary corresponds to a gapless state, exhibiting a critical entanglement scaling $\mathcal{S}_{vN}\propto \ln L$ in the leading order. Even away from the critical line, the critical scaling is observed up to system sizes corresponding to the correlation length in the gapped phase. The critical state provides another important test for the correspondence between the entanglement entropy and the reduced fluctuations. Due to a lack of conservation, the bare spin fluctuations again obey the volume-law scaling. However, as seen in Fig.~\ref{fig:megafigure}(d), the reduced spin fluctuations again follow the same scaling behavior as the entanglement entropy, when scaled with an appropriate parameter-specific factor. Fig.~\ref{fig:megafigure}(e) shows a comparison between the fluctuations of all three spin components. Along the critical line the system is symmetric with respect to rotations around $y$, so that $\delta_r^2S^y\equiv\delta^2S^y$. As expected, the $x$ and $z$ couplings are equal along this line, resulting in the reduced fluctuations of $S^x$ and $S^z$ being identical. 

In Fig~\ref{fig:megafigure}(f) we show the evolution of visibility for all three $S$ operators along a cut in the phase diagram. The particular choice $J_z=0.7J$ helps to highlight the different behavior of visibility, depending on the parameters of the model, and the choice of operators. In general, the more a particular symmetry is broken (rotation around the $x$, $y$, or $z$ axes in the present case), the smaller the value of visibility. Again, as visualized in Fig.~\ref{fig:intro_fig}(a), this stems from the fact that when the symmetry is broken, different spin sectors contribute to the ground state and the total system variance increases, thus reducing the visibility. We can see in Fig.~\ref{fig:megafigure}(f) that when the system is symmetric w. r. t. the rotations around the $z$ axis ($\gamma=0$), $\beta(S^z)=1$ and the visibilities of $S^x$ and $S^y$ are equal. Increasing the value of $\gamma$ and therefore introducing more $S^z$ sectors to the ground state results in a steady decrease of visibility for $S^z$. Similarly, at $\gamma=0.4$, the system becomes symmetric w. r. t. the rotations around the $y$ axis, therefore $\beta(S^y)=1$ and any change from this symmetric point decreases the visibility of the reduced fluctuations of $S^y$. In this region of the phase diagram, the system is never close the $x$ rotational symmetry, so $\beta(S^x)$ remains rather low and does not change significantly.

\emph{Discussion and conclusion}.—The fundamental difficulty of directly measuring entanglement entropy lies in the fact that it is defined in terms of the reduced density matrix, the size of which grows exponentially in the number of degrees of freedom \cite{Cramer2010tomography,Brydges2019probing,Eisert2010PRL}. Thus, even optimized approaches to measuring density matrices exhibit exponential complexity, which fundamentally limits the studies to small systems. However, the entanglement entropy itself is a single coarse-grained characteristic of the full reduced density matrix, and some of its properties could be shared by observables whose measurement does not suffer from exponential complexity \cite{huang2020predicting}. In the present work, we proposed reduced fluctuations of sums of on-site operators as a scalable experimental probe to obtaining the entanglement entropy scaling laws. The general approach was illustrated in interacting spin chains, but is equally applicable to many-body systems of identical particles as well as quantum devices with qubits.  

The complexity associated with measuring the reduced fluctuation with the required relative accuracy is characterized by visibility $\beta$. The standard error of the reduced fluctuation scales as those of typical extensive observables $\Delta(\delta^2_r\hat{A})\sim L^D/\sqrt{\cal N}$, where $L$ is the characteristic linear system size and ${\cal N}$ is the number of repeated measurements of all on-site observables. This polynomial scaling should be contrasted to the exponential scaling of all direct approaches to measure entanglement entropy based on state tomography. While the exponential scaling will always limit the state tomography to a small number of qubits, probing the entanglement scaling through reduced fluctuations opens up an experimental route to study systems from a few dozen to a few hundred qubits, which is the characteristic size of existing and near future quantum computers and quantum simulator systems. Recent advances have revealed that some of the most intriguing realizations of quantum phases of matter are supported by NISQ devices and quantum simulator systems. The method of reduced fluctuations is particularly convenient for studying the entanglement scaling in these systems.

\emph{Acknowledgements}.—T.O. and A.G.M. acknowledge Jane and Aatos Erkko Foundation for financial support. T.O., and  K.P. acknowledges support from the Finnish Research Council through  projects  362573 (T.O.) and 363879 (K.P.). We gratefully acknowledge Polish high-performance computing infrastructure PLGrid (HPC Center: ACK Cyfronet AGH) for providing computer facilities and support within computational grant no. PLG/2024/017795.

\emph{Data availability}.—
The data that support the findings of this article are openly available \cite{zenodo}.
\bibliography{bibliography}

\begin{thebibliography}{50}%
\makeatletter
\providecommand \@ifxundefined [1]{%
 \@ifx{#1\undefined}
}%
\providecommand \@ifnum [1]{%
 \ifnum #1\expandafter \@firstoftwo
 \else \expandafter \@secondoftwo
 \fi
}%
\providecommand \@ifx [1]{%
 \ifx #1\expandafter \@firstoftwo
 \else \expandafter \@secondoftwo
 \fi
}%
\providecommand \natexlab [1]{#1}%
\providecommand \enquote  [1]{``#1''}%
\providecommand \bibnamefont  [1]{#1}%
\providecommand \bibfnamefont [1]{#1}%
\providecommand \citenamefont [1]{#1}%
\providecommand \href@noop [0]{\@secondoftwo}%
\providecommand \href [0]{\begingroup \@sanitize@url \@href}%
\providecommand \@href[1]{\@@startlink{#1}\@@href}%
\providecommand \@@href[1]{\endgroup#1\@@endlink}%
\providecommand \@sanitize@url [0]{\catcode `\\12\catcode `\$12\catcode
  `\&12\catcode `\#12\catcode `\^12\catcode `\_12\catcode `\%12\relax}%
\providecommand \@@startlink[1]{}%
\providecommand \@@endlink[0]{}%
\providecommand \url  [0]{\begingroup\@sanitize@url \@url }%
\providecommand \@url [1]{\endgroup\@href {#1}{\urlprefix }}%
\providecommand \urlprefix  [0]{URL }%
\providecommand \Eprint [0]{\href }%
\providecommand \doibase [0]{https://doi.org/}%
\providecommand \selectlanguage [0]{\@gobble}%
\providecommand \bibinfo  [0]{\@secondoftwo}%
\providecommand \bibfield  [0]{\@secondoftwo}%
\providecommand \translation [1]{[#1]}%
\providecommand \BibitemOpen [0]{}%
\providecommand \bibitemStop [0]{}%
\providecommand \bibitemNoStop [0]{.\EOS\space}%
\providecommand \EOS [0]{\spacefactor3000\relax}%
\providecommand \BibitemShut  [1]{\csname bibitem#1\endcsname}%
\let\auto@bib@innerbib\@empty
\bibitem [{\citenamefont {Wen}(2017)}]{Wen2017}%
  \BibitemOpen
  \bibfield  {author} {\bibinfo {author} {\bibfnamefont {X.-G.}\ \bibnamefont
  {Wen}},\ }\bibfield  {title} {\bibinfo {title} {Colloquium: Zoo of
  quantum-topological phases of matter},\ }\href
  {https://doi.org/10.1103/RevModPhys.89.041004} {\bibfield  {journal}
  {\bibinfo  {journal} {Rev. Mod. Phys.}\ }\textbf {\bibinfo {volume} {89}},\
  \bibinfo {pages} {041004} (\bibinfo {year} {2017})}\BibitemShut {NoStop}%
\bibitem [{\citenamefont {Kitaev}(2009)}]{Kitaev2009}%
  \BibitemOpen
  \bibfield  {author} {\bibinfo {author} {\bibfnamefont {A.}~\bibnamefont
  {Kitaev}},\ }\bibfield  {title} {\bibinfo {title} {Periodic table for
  topological insulators and superconductors},\ }\href
  {https://doi.org/10.1063/1.3149495} {\bibfield  {journal} {\bibinfo
  {journal} {AIP Conf. Proc.}\ }\textbf {\bibinfo {volume} {1134}},\ \bibinfo
  {pages} {22} (\bibinfo {year} {2009})}\BibitemShut {NoStop}%
\bibitem [{\citenamefont {Chiu}\ \emph {et~al.}(2016)\citenamefont {Chiu},
  \citenamefont {Teo}, \citenamefont {Schnyder},\ and\ \citenamefont
  {Ryu}}]{Ryu2016}%
  \BibitemOpen
  \bibfield  {author} {\bibinfo {author} {\bibfnamefont {C.-K.}\ \bibnamefont
  {Chiu}}, \bibinfo {author} {\bibfnamefont {J.~C.~Y.}\ \bibnamefont {Teo}},
  \bibinfo {author} {\bibfnamefont {A.~P.}\ \bibnamefont {Schnyder}},\ and\
  \bibinfo {author} {\bibfnamefont {S.}~\bibnamefont {Ryu}},\ }\bibfield
  {title} {\bibinfo {title} {Classification of topological quantum matter with
  symmetries},\ }\href {https://doi.org/10.1103/RevModPhys.88.035005}
  {\bibfield  {journal} {\bibinfo  {journal} {Rev. Mod. Phys.}\ }\textbf
  {\bibinfo {volume} {88}},\ \bibinfo {pages} {035005} (\bibinfo {year}
  {2016})}\BibitemShut {NoStop}%
\bibitem [{\citenamefont {Horodecki}\ \emph {et~al.}(2009)\citenamefont
  {Horodecki}, \citenamefont {Horodecki}, \citenamefont {Horodecki},\ and\
  \citenamefont {Horodecki}}]{Horodecki}%
  \BibitemOpen
  \bibfield  {author} {\bibinfo {author} {\bibfnamefont {R.}~\bibnamefont
  {Horodecki}}, \bibinfo {author} {\bibfnamefont {P.}~\bibnamefont
  {Horodecki}}, \bibinfo {author} {\bibfnamefont {M.}~\bibnamefont
  {Horodecki}},\ and\ \bibinfo {author} {\bibfnamefont {K.}~\bibnamefont
  {Horodecki}},\ }\bibfield  {title} {\bibinfo {title} {Quantum entanglement},\
  }\href {https://doi.org/10.1103/RevModPhys.81.865} {\bibfield  {journal}
  {\bibinfo  {journal} {Rev. Mod. Phys.}\ }\textbf {\bibinfo {volume} {81}},\
  \bibinfo {pages} {865} (\bibinfo {year} {2009})}\BibitemShut {NoStop}%
\bibitem [{\citenamefont {Zeng}\ \emph {et~al.}(2019)\citenamefont {Zeng},
  \citenamefont {Chen}, \citenamefont {Zhou}, \citenamefont {Wen} \emph
  {et~al.}}]{zeng2019book}%
  \BibitemOpen
  \bibfield  {author} {\bibinfo {author} {\bibfnamefont {B.}~\bibnamefont
  {Zeng}}, \bibinfo {author} {\bibfnamefont {X.}~\bibnamefont {Chen}}, \bibinfo
  {author} {\bibfnamefont {D.-L.}\ \bibnamefont {Zhou}}, \bibinfo {author}
  {\bibfnamefont {X.-G.}\ \bibnamefont {Wen}}, \emph {et~al.},\ }\href@noop {}
  {\emph {\bibinfo {title} {Quantum information meets quantum matter}}}\
  (\bibinfo  {publisher} {Springer},\ \bibinfo {year} {2019})\BibitemShut
  {NoStop}%
\bibitem [{\citenamefont {Amico}\ \emph {et~al.}(2008)\citenamefont {Amico},
  \citenamefont {Fazio}, \citenamefont {Osterloh},\ and\ \citenamefont
  {Vedral}}]{vedral2008rmp}%
  \BibitemOpen
  \bibfield  {author} {\bibinfo {author} {\bibfnamefont {L.}~\bibnamefont
  {Amico}}, \bibinfo {author} {\bibfnamefont {R.}~\bibnamefont {Fazio}},
  \bibinfo {author} {\bibfnamefont {A.}~\bibnamefont {Osterloh}},\ and\
  \bibinfo {author} {\bibfnamefont {V.}~\bibnamefont {Vedral}},\ }\bibfield
  {title} {\bibinfo {title} {Entanglement in many-body systems},\ }\href
  {https://doi.org/10.1103/RevModPhys.80.517} {\bibfield  {journal} {\bibinfo
  {journal} {Rev. Mod. Phys.}\ }\textbf {\bibinfo {volume} {80}},\ \bibinfo
  {pages} {517} (\bibinfo {year} {2008})}\BibitemShut {NoStop}%
\bibitem [{\citenamefont {Plenio}\ and\ \citenamefont
  {Virmani}(2014)}]{plenio2014introduction}%
  \BibitemOpen
  \bibfield  {author} {\bibinfo {author} {\bibfnamefont {M.~B.}\ \bibnamefont
  {Plenio}}\ and\ \bibinfo {author} {\bibfnamefont {S.~S.}\ \bibnamefont
  {Virmani}},\ }\bibinfo {title} {An introduction to entanglement theory},\ in\
  \href {https://doi.org/10.1007/978-3-319-04063-9_8} {\emph {\bibinfo
  {booktitle} {Quantum Information and Coherence}}},\ \bibinfo {editor} {edited
  by\ \bibinfo {editor} {\bibfnamefont {E.}~\bibnamefont {Andersson}}\ and\
  \bibinfo {editor} {\bibfnamefont {P.}~\bibnamefont {{\"O}hberg}}}\ (\bibinfo
  {publisher} {Springer International Publishing},\ \bibinfo {address} {Cham},\
  \bibinfo {year} {2014})\ p.\ \bibinfo {pages} {173}\BibitemShut {NoStop}%
\bibitem [{\citenamefont {Laflorencie}(2016)}]{laflorencie2016quantum}%
  \BibitemOpen
  \bibfield  {author} {\bibinfo {author} {\bibfnamefont {N.}~\bibnamefont
  {Laflorencie}},\ }\bibfield  {title} {\bibinfo {title} {Quantum entanglement
  in condensed matter systems},\ }\href
  {https://doi.org/10.1016/j.physrep.2016.06.008} {\bibfield  {journal}
  {\bibinfo  {journal} {Phys. Rep.}\ }\textbf {\bibinfo {volume} {646}},\
  \bibinfo {pages} {1} (\bibinfo {year} {2016})}\BibitemShut {NoStop}%
\bibitem [{\citenamefont {Kitaev}\ and\ \citenamefont
  {Preskill}(2006)}]{Kitaev2006}%
  \BibitemOpen
  \bibfield  {author} {\bibinfo {author} {\bibfnamefont {A.}~\bibnamefont
  {Kitaev}}\ and\ \bibinfo {author} {\bibfnamefont {J.}~\bibnamefont
  {Preskill}},\ }\bibfield  {title} {\bibinfo {title} {Topological entanglement
  entropy},\ }\href {https://doi.org/10.1103/PhysRevLett.96.110404} {\bibfield
  {journal} {\bibinfo  {journal} {Phys. Rev. Lett.}\ }\textbf {\bibinfo
  {volume} {96}},\ \bibinfo {pages} {110404} (\bibinfo {year}
  {2006})}\BibitemShut {NoStop}%
\bibitem [{\citenamefont {Fidkowski}\ and\ \citenamefont
  {Kitaev}(2011)}]{Fidkowski2011}%
  \BibitemOpen
  \bibfield  {author} {\bibinfo {author} {\bibfnamefont {L.}~\bibnamefont
  {Fidkowski}}\ and\ \bibinfo {author} {\bibfnamefont {A.}~\bibnamefont
  {Kitaev}},\ }\bibfield  {title} {\bibinfo {title} {Topological phases of
  fermions in one dimension},\ }\href
  {https://doi.org/10.1103/PhysRevB.83.075103} {\bibfield  {journal} {\bibinfo
  {journal} {Phys. Rev. B}\ }\textbf {\bibinfo {volume} {83}},\ \bibinfo
  {pages} {075103} (\bibinfo {year} {2011})}\BibitemShut {NoStop}%
\bibitem [{\citenamefont {Calabrese}\ and\ \citenamefont
  {Cardy}(2007)}]{calabrese2007entanglement}%
  \BibitemOpen
  \bibfield  {author} {\bibinfo {author} {\bibfnamefont {P.}~\bibnamefont
  {Calabrese}}\ and\ \bibinfo {author} {\bibfnamefont {J.}~\bibnamefont
  {Cardy}},\ }\bibfield  {title} {\bibinfo {title} {Entanglement and
  correlation functions following a local quench: a conformal field theory
  approach},\ }\href {https://doi.org/10.1088/1742-5468/2007/10/P10004}
  {\bibfield  {journal} {\bibinfo  {journal} {J. Stat. Mech.: Theory Exp.}\
  }\textbf {\bibinfo {volume} {2007}}\bibinfo  {number} { (10)},\ \bibinfo
  {pages} {P10004}}\BibitemShut {NoStop}%
\bibitem [{\citenamefont {Eisert}\ \emph {et~al.}(2015)\citenamefont {Eisert},
  \citenamefont {Friesdorf},\ and\ \citenamefont
  {Gogolin}}]{eisert2015quantum}%
  \BibitemOpen
\bibfield  {number} {  }\bibfield  {author} {\bibinfo {author} {\bibfnamefont
  {J.}~\bibnamefont {Eisert}}, \bibinfo {author} {\bibfnamefont
  {M.}~\bibnamefont {Friesdorf}},\ and\ \bibinfo {author} {\bibfnamefont
  {C.}~\bibnamefont {Gogolin}},\ }\bibfield  {title} {\bibinfo {title} {Quantum
  many-body systems out of equilibrium},\ }\href
  {https://doi.org/10.1038/nphys3215} {\bibfield  {journal} {\bibinfo
  {journal} {Nat. Phys.}\ }\textbf {\bibinfo {volume} {11}},\ \bibinfo {pages}
  {124} (\bibinfo {year} {2015})}\BibitemShut {NoStop}%
\bibitem [{\citenamefont {Osterloh}\ \emph {et~al.}(2002)\citenamefont
  {Osterloh}, \citenamefont {Amico}, \citenamefont {Falci},\ and\ \citenamefont
  {Fazio}}]{osterloh2002scaling}%
  \BibitemOpen
  \bibfield  {author} {\bibinfo {author} {\bibfnamefont {A.}~\bibnamefont
  {Osterloh}}, \bibinfo {author} {\bibfnamefont {L.}~\bibnamefont {Amico}},
  \bibinfo {author} {\bibfnamefont {G.}~\bibnamefont {Falci}},\ and\ \bibinfo
  {author} {\bibfnamefont {R.}~\bibnamefont {Fazio}},\ }\bibfield  {title}
  {\bibinfo {title} {Scaling of entanglement close to a quantum phase
  transition},\ }\href {https://doi.org/10.1038/416608a} {\bibfield  {journal}
  {\bibinfo  {journal} {Nature}\ }\textbf {\bibinfo {volume} {416}},\ \bibinfo
  {pages} {608} (\bibinfo {year} {2002})}\BibitemShut {NoStop}%
\bibitem [{\citenamefont {Vidal}\ \emph {et~al.}(2003)\citenamefont {Vidal},
  \citenamefont {Latorre}, \citenamefont {Rico},\ and\ \citenamefont
  {Kitaev}}]{Kitaev2003}%
  \BibitemOpen
  \bibfield  {author} {\bibinfo {author} {\bibfnamefont {G.}~\bibnamefont
  {Vidal}}, \bibinfo {author} {\bibfnamefont {J.~I.}\ \bibnamefont {Latorre}},
  \bibinfo {author} {\bibfnamefont {E.}~\bibnamefont {Rico}},\ and\ \bibinfo
  {author} {\bibfnamefont {A.}~\bibnamefont {Kitaev}},\ }\bibfield  {title}
  {\bibinfo {title} {Entanglement in quantum critical phenomena},\ }\href
  {https://doi.org/10.1103/PhysRevLett.90.227902} {\bibfield  {journal}
  {\bibinfo  {journal} {Phys. Rev. Lett.}\ }\textbf {\bibinfo {volume} {90}},\
  \bibinfo {pages} {227902} (\bibinfo {year} {2003})}\BibitemShut {NoStop}%
\bibitem [{\citenamefont {Hastings}(2007)}]{hastings2007area}%
  \BibitemOpen
  \bibfield  {author} {\bibinfo {author} {\bibfnamefont {M.~B.}\ \bibnamefont
  {Hastings}},\ }\bibfield  {title} {\bibinfo {title} {An area law for
  one-dimensional quantum systems},\ }\href
  {https://doi.org/10.1088/1742-5468/2007/08/P08024} {\bibfield  {journal}
  {\bibinfo  {journal} {J. Stat. Mech.: Theory Exp.}\ }\textbf {\bibinfo
  {volume} {2007}}\bibinfo  {number} { (08)},\ \bibinfo {pages}
  {P08024}}\BibitemShut {NoStop}%
\bibitem [{\citenamefont {Plenio}\ \emph {et~al.}(2005)\citenamefont {Plenio},
  \citenamefont {Eisert}, \citenamefont {Drei\ss{}ig},\ and\ \citenamefont
  {Cramer}}]{Plenio2005}%
  \BibitemOpen
\bibfield  {number} {  }\bibfield  {author} {\bibinfo {author} {\bibfnamefont
  {M.~B.}\ \bibnamefont {Plenio}}, \bibinfo {author} {\bibfnamefont
  {J.}~\bibnamefont {Eisert}}, \bibinfo {author} {\bibfnamefont
  {J.}~\bibnamefont {Drei\ss{}ig}},\ and\ \bibinfo {author} {\bibfnamefont
  {M.}~\bibnamefont {Cramer}},\ }\bibfield  {title} {\bibinfo {title} {Entropy,
  entanglement, and area: Analytical results for harmonic lattice systems},\
  }\href {https://doi.org/10.1103/PhysRevLett.94.060503} {\bibfield  {journal}
  {\bibinfo  {journal} {Phys. Rev. Lett.}\ }\textbf {\bibinfo {volume} {94}},\
  \bibinfo {pages} {060503} (\bibinfo {year} {2005})}\BibitemShut {NoStop}%
\bibitem [{\citenamefont {Wolf}(2006)}]{Wolf2006}%
  \BibitemOpen
  \bibfield  {author} {\bibinfo {author} {\bibfnamefont {M.~M.}\ \bibnamefont
  {Wolf}},\ }\bibfield  {title} {\bibinfo {title} {Violation of the entropic
  area law for fermions},\ }\href
  {https://doi.org/10.1103/PhysRevLett.96.010404} {\bibfield  {journal}
  {\bibinfo  {journal} {Phys. Rev. Lett.}\ }\textbf {\bibinfo {volume} {96}},\
  \bibinfo {pages} {010404} (\bibinfo {year} {2006})}\BibitemShut {NoStop}%
\bibitem [{\citenamefont {Levin}\ and\ \citenamefont {Wen}(2006)}]{wen2006}%
  \BibitemOpen
  \bibfield  {author} {\bibinfo {author} {\bibfnamefont {M.}~\bibnamefont
  {Levin}}\ and\ \bibinfo {author} {\bibfnamefont {X.-G.}\ \bibnamefont
  {Wen}},\ }\bibfield  {title} {\bibinfo {title} {Detecting topological order
  in a ground state wave function},\ }\href
  {https://doi.org/10.1103/PhysRevLett.96.110405} {\bibfield  {journal}
  {\bibinfo  {journal} {Phys. Rev. Lett.}\ }\textbf {\bibinfo {volume} {96}},\
  \bibinfo {pages} {110405} (\bibinfo {year} {2006})}\BibitemShut {NoStop}%
\bibitem [{\citenamefont {Jiang}\ \emph {et~al.}(2012)\citenamefont {Jiang},
  \citenamefont {Wang},\ and\ \citenamefont
  {Balents}}]{Balents2012identifying}%
  \BibitemOpen
  \bibfield  {author} {\bibinfo {author} {\bibfnamefont {H.-C.}\ \bibnamefont
  {Jiang}}, \bibinfo {author} {\bibfnamefont {Z.}~\bibnamefont {Wang}},\ and\
  \bibinfo {author} {\bibfnamefont {L.}~\bibnamefont {Balents}},\ }\bibfield
  {title} {\bibinfo {title} {Identifying topological order by entanglement
  entropy},\ }\href {https://doi.org/10.1038/nphys2465} {\bibfield  {journal}
  {\bibinfo  {journal} {Nat. Phys.}\ }\textbf {\bibinfo {volume} {8}},\
  \bibinfo {pages} {902} (\bibinfo {year} {2012})}\BibitemShut {NoStop}%
\bibitem [{\citenamefont {Li}\ \emph {et~al.}(2018)\citenamefont {Li},
  \citenamefont {Chen},\ and\ \citenamefont {Fisher}}]{Fisher2018}%
  \BibitemOpen
  \bibfield  {author} {\bibinfo {author} {\bibfnamefont {Y.}~\bibnamefont
  {Li}}, \bibinfo {author} {\bibfnamefont {X.}~\bibnamefont {Chen}},\ and\
  \bibinfo {author} {\bibfnamefont {M.~P.~A.}\ \bibnamefont {Fisher}},\
  }\bibfield  {title} {\bibinfo {title} {Quantum zeno effect and the many-body
  entanglement transition},\ }\href
  {https://doi.org/10.1103/PhysRevB.98.205136} {\bibfield  {journal} {\bibinfo
  {journal} {Phys. Rev. B}\ }\textbf {\bibinfo {volume} {98}},\ \bibinfo
  {pages} {205136} (\bibinfo {year} {2018})}\BibitemShut {NoStop}%
\bibitem [{\citenamefont {Skinner}\ \emph {et~al.}(2019)\citenamefont
  {Skinner}, \citenamefont {Ruhman},\ and\ \citenamefont
  {Nahum}}]{Skinner2019}%
  \BibitemOpen
  \bibfield  {author} {\bibinfo {author} {\bibfnamefont {B.}~\bibnamefont
  {Skinner}}, \bibinfo {author} {\bibfnamefont {J.}~\bibnamefont {Ruhman}},\
  and\ \bibinfo {author} {\bibfnamefont {A.}~\bibnamefont {Nahum}},\ }\bibfield
   {title} {\bibinfo {title} {Measurement-induced phase transitions in the
  dynamics of entanglement},\ }\href
  {https://doi.org/10.1103/PhysRevX.9.031009} {\bibfield  {journal} {\bibinfo
  {journal} {Phys. Rev. X}\ }\textbf {\bibinfo {volume} {9}},\ \bibinfo {pages}
  {031009} (\bibinfo {year} {2019})}\BibitemShut {NoStop}%
\bibitem [{\citenamefont {Fisher}\ \emph {et~al.}(2023)\citenamefont {Fisher},
  \citenamefont {Khemani}, \citenamefont {Nahum},\ and\ \citenamefont
  {Vijay}}]{Fisher2022review}%
  \BibitemOpen
  \bibfield  {author} {\bibinfo {author} {\bibfnamefont {M.~P.}\ \bibnamefont
  {Fisher}}, \bibinfo {author} {\bibfnamefont {V.}~\bibnamefont {Khemani}},
  \bibinfo {author} {\bibfnamefont {A.}~\bibnamefont {Nahum}},\ and\ \bibinfo
  {author} {\bibfnamefont {S.}~\bibnamefont {Vijay}},\ }\bibfield  {title}
  {\bibinfo {title} {Random quantum circuits},\ }\href
  {https://doi.org/10.1146/annurev-conmatphys-031720-030658} {\bibfield
  {journal} {\bibinfo  {journal} {Annu. Rev. Condens. Matter Phys.}\ }\textbf
  {\bibinfo {volume} {14}},\ \bibinfo {pages} {null} (\bibinfo {year}
  {2023})}\BibitemShut {NoStop}%
\bibitem [{\citenamefont {Hamma}\ \emph {et~al.}(2005)\citenamefont {Hamma},
  \citenamefont {Ionicioiu},\ and\ \citenamefont {Zanardi}}]{Zanardi2005}%
  \BibitemOpen
  \bibfield  {author} {\bibinfo {author} {\bibfnamefont {A.}~\bibnamefont
  {Hamma}}, \bibinfo {author} {\bibfnamefont {R.}~\bibnamefont {Ionicioiu}},\
  and\ \bibinfo {author} {\bibfnamefont {P.}~\bibnamefont {Zanardi}},\
  }\bibfield  {title} {\bibinfo {title} {Bipartite entanglement and entropic
  boundary law in lattice spin systems},\ }\href
  {https://doi.org/10.1103/PhysRevA.71.022315} {\bibfield  {journal} {\bibinfo
  {journal} {Phys. Rev. A}\ }\textbf {\bibinfo {volume} {71}},\ \bibinfo
  {pages} {022315} (\bibinfo {year} {2005})}\BibitemShut {NoStop}%
\bibitem [{\citenamefont {Swingle}(2010)}]{Swingle2010}%
  \BibitemOpen
  \bibfield  {author} {\bibinfo {author} {\bibfnamefont {B.}~\bibnamefont
  {Swingle}},\ }\bibfield  {title} {\bibinfo {title} {{Entanglement Entropy and
  the Fermi Surface}},\ }\href {https://doi.org/10.1103/PhysRevLett.105.050502}
  {\bibfield  {journal} {\bibinfo  {journal} {Phys. Rev. Lett.}\ }\textbf
  {\bibinfo {volume} {105}},\ \bibinfo {pages} {050502} (\bibinfo {year}
  {2010})}\BibitemShut {NoStop}%
\bibitem [{\citenamefont {Bauer}\ and\ \citenamefont
  {Nayak}(2013)}]{Nayak2013}%
  \BibitemOpen
  \bibfield  {author} {\bibinfo {author} {\bibfnamefont {B.}~\bibnamefont
  {Bauer}}\ and\ \bibinfo {author} {\bibfnamefont {C.}~\bibnamefont {Nayak}},\
  }\bibfield  {title} {\bibinfo {title} {Area laws in a many-body localized
  state and its implications for topological order},\ }\href
  {https://doi.org/10.1088/1742-5468/2013/09/p09005} {\bibfield  {journal}
  {\bibinfo  {journal} {J. Stat. Mech.: Theory Exp.}\ }\textbf {\bibinfo
  {volume} {2013}}\bibinfo  {number} { (09)},\ \bibinfo {pages}
  {P09005}}\BibitemShut {NoStop}%
\bibitem [{\citenamefont {Abanin}\ \emph {et~al.}(2019)\citenamefont {Abanin},
  \citenamefont {Altman}, \citenamefont {Bloch},\ and\ \citenamefont
  {Serbyn}}]{Abanin2019rmp}%
  \BibitemOpen
\bibfield  {number} {  }\bibfield  {author} {\bibinfo {author} {\bibfnamefont
  {D.~A.}\ \bibnamefont {Abanin}}, \bibinfo {author} {\bibfnamefont
  {E.}~\bibnamefont {Altman}}, \bibinfo {author} {\bibfnamefont
  {I.}~\bibnamefont {Bloch}},\ and\ \bibinfo {author} {\bibfnamefont
  {M.}~\bibnamefont {Serbyn}},\ }\bibfield  {title} {\bibinfo {title}
  {Colloquium: Many-body localization, thermalization, and entanglement},\
  }\href {https://doi.org/10.1103/RevModPhys.91.021001} {\bibfield  {journal}
  {\bibinfo  {journal} {Rev. Mod. Phys.}\ }\textbf {\bibinfo {volume} {91}},\
  \bibinfo {pages} {021001} (\bibinfo {year} {2019})}\BibitemShut {NoStop}%
\bibitem [{\citenamefont {Cramer}\ \emph {et~al.}(2010)\citenamefont {Cramer},
  \citenamefont {Plenio}, \citenamefont {Flammia}, \citenamefont {Somma},
  \citenamefont {Gross}, \citenamefont {Bartlett}, \citenamefont
  {Landon-Cardinal}, \citenamefont {Poulin},\ and\ \citenamefont
  {Liu}}]{Cramer2010tomography}%
  \BibitemOpen
  \bibfield  {author} {\bibinfo {author} {\bibfnamefont {M.}~\bibnamefont
  {Cramer}}, \bibinfo {author} {\bibfnamefont {M.~B.}\ \bibnamefont {Plenio}},
  \bibinfo {author} {\bibfnamefont {S.~T.}\ \bibnamefont {Flammia}}, \bibinfo
  {author} {\bibfnamefont {R.}~\bibnamefont {Somma}}, \bibinfo {author}
  {\bibfnamefont {D.}~\bibnamefont {Gross}}, \bibinfo {author} {\bibfnamefont
  {S.~D.}\ \bibnamefont {Bartlett}}, \bibinfo {author} {\bibfnamefont
  {O.}~\bibnamefont {Landon-Cardinal}}, \bibinfo {author} {\bibfnamefont
  {D.}~\bibnamefont {Poulin}},\ and\ \bibinfo {author} {\bibfnamefont {Y.-K.}\
  \bibnamefont {Liu}},\ }\bibfield  {title} {\bibinfo {title} {Efficient
  quantum state tomography},\ }\href {https://doi.org/10.1038/ncomms1147}
  {\bibfield  {journal} {\bibinfo  {journal} {Nat. Commun.}\ }\textbf {\bibinfo
  {volume} {1}},\ \bibinfo {pages} {149} (\bibinfo {year} {2010})}\BibitemShut
  {NoStop}%
\bibitem [{\citenamefont {Eisert}\ \emph {et~al.}(2010)\citenamefont {Eisert},
  \citenamefont {Cramer},\ and\ \citenamefont {Plenio}}]{Eisert2010}%
  \BibitemOpen
  \bibfield  {author} {\bibinfo {author} {\bibfnamefont {J.}~\bibnamefont
  {Eisert}}, \bibinfo {author} {\bibfnamefont {M.}~\bibnamefont {Cramer}},\
  and\ \bibinfo {author} {\bibfnamefont {M.~B.}\ \bibnamefont {Plenio}},\
  }\bibfield  {title} {\bibinfo {title} {Colloquium: Area laws for the
  entanglement entropy},\ }\href {https://doi.org/10.1103/RevModPhys.82.277}
  {\bibfield  {journal} {\bibinfo  {journal} {Rev. Mod. Phys.}\ }\textbf
  {\bibinfo {volume} {82}},\ \bibinfo {pages} {277} (\bibinfo {year}
  {2010})}\BibitemShut {NoStop}%
\bibitem [{\citenamefont {Islam}\ \emph {et~al.}(2015)\citenamefont {Islam},
  \citenamefont {Ma}, \citenamefont {Preiss}, \citenamefont {Tai},
  \citenamefont {Lukin}, \citenamefont {Rispoli},\ and\ \citenamefont
  {Greiner}}]{Greiner2015measuring}%
  \BibitemOpen
  \bibfield  {author} {\bibinfo {author} {\bibfnamefont {R.}~\bibnamefont
  {Islam}}, \bibinfo {author} {\bibfnamefont {R.}~\bibnamefont {Ma}}, \bibinfo
  {author} {\bibfnamefont {P.~M.}\ \bibnamefont {Preiss}}, \bibinfo {author}
  {\bibfnamefont {M.~E.}\ \bibnamefont {Tai}}, \bibinfo {author} {\bibfnamefont
  {A.}~\bibnamefont {Lukin}}, \bibinfo {author} {\bibfnamefont
  {M.}~\bibnamefont {Rispoli}},\ and\ \bibinfo {author} {\bibfnamefont
  {M.}~\bibnamefont {Greiner}},\ }\bibfield  {title} {\bibinfo {title}
  {Measuring entanglement entropy in a quantum many-body system},\ }\href
  {https://doi.org/10.1038/nature15750} {\bibfield  {journal} {\bibinfo
  {journal} {Nature}\ }\textbf {\bibinfo {volume} {528}},\ \bibinfo {pages}
  {77} (\bibinfo {year} {2015})}\BibitemShut {NoStop}%
\bibitem [{\citenamefont {Brydges}\ \emph {et~al.}(2019)\citenamefont
  {Brydges}, \citenamefont {Elben}, \citenamefont {Jurcevic}, \citenamefont
  {Vermersch}, \citenamefont {Maier}, \citenamefont {Lanyon}, \citenamefont
  {Zoller}, \citenamefont {Blatt},\ and\ \citenamefont
  {Roos}}]{Brydges2019probing}%
  \BibitemOpen
  \bibfield  {author} {\bibinfo {author} {\bibfnamefont {T.}~\bibnamefont
  {Brydges}}, \bibinfo {author} {\bibfnamefont {A.}~\bibnamefont {Elben}},
  \bibinfo {author} {\bibfnamefont {P.}~\bibnamefont {Jurcevic}}, \bibinfo
  {author} {\bibfnamefont {B.}~\bibnamefont {Vermersch}}, \bibinfo {author}
  {\bibfnamefont {C.}~\bibnamefont {Maier}}, \bibinfo {author} {\bibfnamefont
  {B.~P.}\ \bibnamefont {Lanyon}}, \bibinfo {author} {\bibfnamefont
  {P.}~\bibnamefont {Zoller}}, \bibinfo {author} {\bibfnamefont
  {R.}~\bibnamefont {Blatt}},\ and\ \bibinfo {author} {\bibfnamefont {C.~F.}\
  \bibnamefont {Roos}},\ }\bibfield  {title} {\bibinfo {title} {Probing
  r{\'e}nyi entanglement entropy via randomized measurements},\ }\href
  {https://doi.org/10.1126/science.aau496} {\bibfield  {journal} {\bibinfo
  {journal} {Science}\ }\textbf {\bibinfo {volume} {364}},\ \bibinfo {pages}
  {260} (\bibinfo {year} {2019})}\BibitemShut {NoStop}%
\bibitem [{\citenamefont {Wolf}\ \emph {et~al.}(2008)\citenamefont {Wolf},
  \citenamefont {Verstraete}, \citenamefont {Hastings},\ and\ \citenamefont
  {Cirac}}]{Wolf2007}%
  \BibitemOpen
  \bibfield  {author} {\bibinfo {author} {\bibfnamefont {M.~M.}\ \bibnamefont
  {Wolf}}, \bibinfo {author} {\bibfnamefont {F.}~\bibnamefont {Verstraete}},
  \bibinfo {author} {\bibfnamefont {M.~B.}\ \bibnamefont {Hastings}},\ and\
  \bibinfo {author} {\bibfnamefont {J.~I.}\ \bibnamefont {Cirac}},\ }\bibfield
  {title} {\bibinfo {title} {Area laws in quantum systems: Mutual information
  and correlations},\ }\href {https://doi.org/10.1103/PhysRevLett.100.070502}
  {\bibfield  {journal} {\bibinfo  {journal} {Phys. Rev. Lett.}\ }\textbf
  {\bibinfo {volume} {100}},\ \bibinfo {pages} {070502} (\bibinfo {year}
  {2008})}\BibitemShut {NoStop}%
\bibitem [{\citenamefont {Verstraete}\ \emph {et~al.}(2004)\citenamefont
  {Verstraete}, \citenamefont {Popp},\ and\ \citenamefont
  {Cirac}}]{Verstraete2004}%
  \BibitemOpen
  \bibfield  {author} {\bibinfo {author} {\bibfnamefont {F.}~\bibnamefont
  {Verstraete}}, \bibinfo {author} {\bibfnamefont {M.}~\bibnamefont {Popp}},\
  and\ \bibinfo {author} {\bibfnamefont {J.~I.}\ \bibnamefont {Cirac}},\
  }\bibfield  {title} {\bibinfo {title} {Entanglement versus correlations in
  spin systems},\ }\href {https://doi.org/10.1103/PhysRevLett.92.027901}
  {\bibfield  {journal} {\bibinfo  {journal} {Phys. Rev. Lett.}\ }\textbf
  {\bibinfo {volume} {92}},\ \bibinfo {pages} {027901} (\bibinfo {year}
  {2004})}\BibitemShut {NoStop}%
\bibitem [{\citenamefont {Klich}(2006)}]{Klich2006}%
  \BibitemOpen
  \bibfield  {author} {\bibinfo {author} {\bibfnamefont {I.}~\bibnamefont
  {Klich}},\ }\bibfield  {title} {\bibinfo {title} {Lower entropy bounds and
  particle number fluctuations in a fermi sea},\ }\href
  {https://doi.org/10.1088/0305-4470/39/4/l02} {\bibfield  {journal} {\bibinfo
  {journal} {Journal of Physics A: Mathematical and General}\ }\textbf
  {\bibinfo {volume} {39}},\ \bibinfo {pages} {L85} (\bibinfo {year}
  {2006})}\BibitemShut {NoStop}%
\bibitem [{\citenamefont {Klich}\ and\ \citenamefont
  {Levitov}(2009)}]{Klich2009}%
  \BibitemOpen
  \bibfield  {author} {\bibinfo {author} {\bibfnamefont {I.}~\bibnamefont
  {Klich}}\ and\ \bibinfo {author} {\bibfnamefont {L.}~\bibnamefont
  {Levitov}},\ }\bibfield  {title} {\bibinfo {title} {Quantum noise as an
  entanglement meter},\ }\href {https://doi.org/10.1103/PhysRevLett.102.100502}
  {\bibfield  {journal} {\bibinfo  {journal} {Phys. Rev. Lett.}\ }\textbf
  {\bibinfo {volume} {102}},\ \bibinfo {pages} {100502} (\bibinfo {year}
  {2009})}\BibitemShut {NoStop}%
\bibitem [{\citenamefont {Song}\ \emph {et~al.}(2010)\citenamefont {Song},
  \citenamefont {Rachel},\ and\ \citenamefont {Le~Hur}}]{LeHur2010}%
  \BibitemOpen
  \bibfield  {author} {\bibinfo {author} {\bibfnamefont {H.~F.}\ \bibnamefont
  {Song}}, \bibinfo {author} {\bibfnamefont {S.}~\bibnamefont {Rachel}},\ and\
  \bibinfo {author} {\bibfnamefont {K.}~\bibnamefont {Le~Hur}},\ }\bibfield
  {title} {\bibinfo {title} {General relation between entanglement and
  fluctuations in one dimension},\ }\href
  {https://doi.org/10.1103/PhysRevB.82.012405} {\bibfield  {journal} {\bibinfo
  {journal} {Phys. Rev. B}\ }\textbf {\bibinfo {volume} {82}},\ \bibinfo
  {pages} {012405} (\bibinfo {year} {2010})}\BibitemShut {NoStop}%
\bibitem [{\citenamefont {Song}\ \emph {et~al.}(2012)\citenamefont {Song},
  \citenamefont {Rachel}, \citenamefont {Flindt}, \citenamefont {Klich},
  \citenamefont {Laflorencie},\ and\ \citenamefont {Le~Hur}}]{LeHur2012}%
  \BibitemOpen
  \bibfield  {author} {\bibinfo {author} {\bibfnamefont {H.~F.}\ \bibnamefont
  {Song}}, \bibinfo {author} {\bibfnamefont {S.}~\bibnamefont {Rachel}},
  \bibinfo {author} {\bibfnamefont {C.}~\bibnamefont {Flindt}}, \bibinfo
  {author} {\bibfnamefont {I.}~\bibnamefont {Klich}}, \bibinfo {author}
  {\bibfnamefont {N.}~\bibnamefont {Laflorencie}},\ and\ \bibinfo {author}
  {\bibfnamefont {K.}~\bibnamefont {Le~Hur}},\ }\bibfield  {title} {\bibinfo
  {title} {Bipartite fluctuations as a probe of many-body entanglement},\
  }\href {https://doi.org/10.1103/PhysRevB.85.035409} {\bibfield  {journal}
  {\bibinfo  {journal} {Phys. Rev. B}\ }\textbf {\bibinfo {volume} {85}},\
  \bibinfo {pages} {035409} (\bibinfo {year} {2012})}\BibitemShut {NoStop}%
\bibitem [{\citenamefont {P\"oyh\"onen}\ \emph {et~al.}(2022)\citenamefont
  {P\"oyh\"onen}, \citenamefont {Moghaddam},\ and\ \citenamefont
  {Ojanen}}]{Poyhonen2022}%
  \BibitemOpen
  \bibfield  {author} {\bibinfo {author} {\bibfnamefont {K.}~\bibnamefont
  {P\"oyh\"onen}}, \bibinfo {author} {\bibfnamefont {A.~G.}\ \bibnamefont
  {Moghaddam}},\ and\ \bibinfo {author} {\bibfnamefont {T.}~\bibnamefont
  {Ojanen}},\ }\bibfield  {title} {\bibinfo {title} {Many-body entanglement and
  topology from uncertainties and measurement-induced modes},\ }\href
  {https://doi.org/10.1103/PhysRevResearch.4.023200} {\bibfield  {journal}
  {\bibinfo  {journal} {Phys. Rev. Res.}\ }\textbf {\bibinfo {volume} {4}},\
  \bibinfo {pages} {023200} (\bibinfo {year} {2022})}\BibitemShut {NoStop}%
\bibitem [{\citenamefont {White}(1992)}]{White1992}%
  \BibitemOpen
  \bibfield  {author} {\bibinfo {author} {\bibfnamefont {S.~R.}\ \bibnamefont
  {White}},\ }\bibfield  {title} {\bibinfo {title} {Density matrix formulation
  for quantum renormalization groups},\ }\href
  {https://doi.org/10.1103/PhysRevLett.69.2863} {\bibfield  {journal} {\bibinfo
   {journal} {Phys. Rev. Lett.}\ }\textbf {\bibinfo {volume} {69}},\ \bibinfo
  {pages} {2863} (\bibinfo {year} {1992})}\BibitemShut {NoStop}%
\bibitem [{\citenamefont {Schollw\"ock}(2005)}]{Schollwock2005dmrg}%
  \BibitemOpen
  \bibfield  {author} {\bibinfo {author} {\bibfnamefont {U.}~\bibnamefont
  {Schollw\"ock}},\ }\bibfield  {title} {\bibinfo {title} {The density-matrix
  renormalization group},\ }\href {https://doi.org/10.1103/RevModPhys.77.259}
  {\bibfield  {journal} {\bibinfo  {journal} {Rev. Mod. Phys.}\ }\textbf
  {\bibinfo {volume} {77}},\ \bibinfo {pages} {259} (\bibinfo {year}
  {2005})}\BibitemShut {NoStop}%
\bibitem [{\citenamefont {Verstraete}\ \emph {et~al.}(2023)\citenamefont
  {Verstraete}, \citenamefont {Nishino}, \citenamefont {Schollw{\"o}ck},
  \citenamefont {Ba{\~n}uls}, \citenamefont {Chan},\ and\ \citenamefont
  {Stoudenmire}}]{verstraete2023}%
  \BibitemOpen
  \bibfield  {author} {\bibinfo {author} {\bibfnamefont {F.}~\bibnamefont
  {Verstraete}}, \bibinfo {author} {\bibfnamefont {T.}~\bibnamefont {Nishino}},
  \bibinfo {author} {\bibfnamefont {U.}~\bibnamefont {Schollw{\"o}ck}},
  \bibinfo {author} {\bibfnamefont {M.~C.}\ \bibnamefont {Ba{\~n}uls}},
  \bibinfo {author} {\bibfnamefont {G.~K.}\ \bibnamefont {Chan}},\ and\
  \bibinfo {author} {\bibfnamefont {M.~E.}\ \bibnamefont {Stoudenmire}},\
  }\bibfield  {title} {\bibinfo {title} {Density matrix renormalization group,
  30 years on},\ }\href@noop {} {\bibfield  {journal} {\bibinfo  {journal}
  {Nature Reviews Physics}\ }\textbf {\bibinfo {volume} {5}},\ \bibinfo {pages}
  {273} (\bibinfo {year} {2023})}\BibitemShut {NoStop}%
\bibitem [{\citenamefont {Preskill}(2018)}]{preskill2018NISQ}%
  \BibitemOpen
  \bibfield  {author} {\bibinfo {author} {\bibfnamefont {J.}~\bibnamefont
  {Preskill}},\ }\bibfield  {title} {\bibinfo {title} {Quantum computing in the
  nisq era and beyond},\ }\href
  {https://doi.org/https://doi.org/10.22331/q-2018-08-06-79} {\bibfield
  {journal} {\bibinfo  {journal} {Quantum}\ }\textbf {\bibinfo {volume} {2}},\
  \bibinfo {pages} {79} (\bibinfo {year} {2018})}\BibitemShut {NoStop}%
\bibitem [{\citenamefont {Altman}\ \emph {et~al.}(2021)\citenamefont {Altman},
  \citenamefont {Brown}, \citenamefont {Carleo}, \citenamefont {Carr},
  \citenamefont {Demler}, \citenamefont {Chin}, \citenamefont {DeMarco},
  \citenamefont {Economou}, \citenamefont {Eriksson}, \citenamefont {Fu},
  \citenamefont {Greiner}, \citenamefont {Hazzard}, \citenamefont {Hulet},
  \citenamefont {Koll\'ar}, \citenamefont {Lev}, \citenamefont {Lukin},
  \citenamefont {Ma}, \citenamefont {Mi}, \citenamefont {Misra}, \citenamefont
  {Monroe}, \citenamefont {Murch}, \citenamefont {Nazario}, \citenamefont {Ni},
  \citenamefont {Potter}, \citenamefont {Roushan}, \citenamefont {Saffman},
  \citenamefont {Schleier-Smith}, \citenamefont {Siddiqi}, \citenamefont
  {Simmonds}, \citenamefont {Singh}, \citenamefont {Spielman}, \citenamefont
  {Temme}, \citenamefont {Weiss}, \citenamefont {Vu\ifmmode \check{c}\else
  \v{c}\fi{}kovi\ifmmode~\acute{c}\else \'{c}\fi{}}, \citenamefont
  {Vuleti\ifmmode~\acute{c}\else \'{c}\fi{}}, \citenamefont {Ye},\ and\
  \citenamefont {Zwierlein}}]{Altman2021}%
  \BibitemOpen
  \bibfield  {author} {\bibinfo {author} {\bibfnamefont {E.}~\bibnamefont
  {Altman}}, \bibinfo {author} {\bibfnamefont {K.~R.}\ \bibnamefont {Brown}},
  \bibinfo {author} {\bibfnamefont {G.}~\bibnamefont {Carleo}}, \bibinfo
  {author} {\bibfnamefont {L.~D.}\ \bibnamefont {Carr}}, \bibinfo {author}
  {\bibfnamefont {E.}~\bibnamefont {Demler}}, \bibinfo {author} {\bibfnamefont
  {C.}~\bibnamefont {Chin}}, \bibinfo {author} {\bibfnamefont {B.}~\bibnamefont
  {DeMarco}}, \bibinfo {author} {\bibfnamefont {S.~E.}\ \bibnamefont
  {Economou}}, \bibinfo {author} {\bibfnamefont {M.~A.}\ \bibnamefont
  {Eriksson}}, \bibinfo {author} {\bibfnamefont {K.-M.~C.}\ \bibnamefont {Fu}},
  \bibinfo {author} {\bibfnamefont {M.}~\bibnamefont {Greiner}}, \bibinfo
  {author} {\bibfnamefont {K.~R.}\ \bibnamefont {Hazzard}}, \bibinfo {author}
  {\bibfnamefont {R.~G.}\ \bibnamefont {Hulet}}, \bibinfo {author}
  {\bibfnamefont {A.~J.}\ \bibnamefont {Koll\'ar}}, \bibinfo {author}
  {\bibfnamefont {B.~L.}\ \bibnamefont {Lev}}, \bibinfo {author} {\bibfnamefont
  {M.~D.}\ \bibnamefont {Lukin}}, \bibinfo {author} {\bibfnamefont
  {R.}~\bibnamefont {Ma}}, \bibinfo {author} {\bibfnamefont {X.}~\bibnamefont
  {Mi}}, \bibinfo {author} {\bibfnamefont {S.}~\bibnamefont {Misra}}, \bibinfo
  {author} {\bibfnamefont {C.}~\bibnamefont {Monroe}}, \bibinfo {author}
  {\bibfnamefont {K.}~\bibnamefont {Murch}}, \bibinfo {author} {\bibfnamefont
  {Z.}~\bibnamefont {Nazario}}, \bibinfo {author} {\bibfnamefont {K.-K.}\
  \bibnamefont {Ni}}, \bibinfo {author} {\bibfnamefont {A.~C.}\ \bibnamefont
  {Potter}}, \bibinfo {author} {\bibfnamefont {P.}~\bibnamefont {Roushan}},
  \bibinfo {author} {\bibfnamefont {M.}~\bibnamefont {Saffman}}, \bibinfo
  {author} {\bibfnamefont {M.}~\bibnamefont {Schleier-Smith}}, \bibinfo
  {author} {\bibfnamefont {I.}~\bibnamefont {Siddiqi}}, \bibinfo {author}
  {\bibfnamefont {R.}~\bibnamefont {Simmonds}}, \bibinfo {author}
  {\bibfnamefont {M.}~\bibnamefont {Singh}}, \bibinfo {author} {\bibfnamefont
  {I.}~\bibnamefont {Spielman}}, \bibinfo {author} {\bibfnamefont
  {K.}~\bibnamefont {Temme}}, \bibinfo {author} {\bibfnamefont {D.~S.}\
  \bibnamefont {Weiss}}, \bibinfo {author} {\bibfnamefont {J.}~\bibnamefont
  {Vu\ifmmode \check{c}\else \v{c}\fi{}kovi\ifmmode~\acute{c}\else
  \'{c}\fi{}}}, \bibinfo {author} {\bibfnamefont {V.}~\bibnamefont
  {Vuleti\ifmmode~\acute{c}\else \'{c}\fi{}}}, \bibinfo {author} {\bibfnamefont
  {J.}~\bibnamefont {Ye}},\ and\ \bibinfo {author} {\bibfnamefont
  {M.}~\bibnamefont {Zwierlein}},\ }\bibfield  {title} {\bibinfo {title}
  {Quantum simulators: Architectures and opportunities},\ }\href
  {https://doi.org/10.1103/PRXQuantum.2.017003} {\bibfield  {journal} {\bibinfo
   {journal} {PRX Quantum}\ }\textbf {\bibinfo {volume} {2}},\ \bibinfo {pages}
  {017003} (\bibinfo {year} {2021})}\BibitemShut {NoStop}%
\bibitem [{\citenamefont {Hoke~\textit{et al.,} {(Google Quantum AI and
  Collaborators)}}(2023)}]{google2023measurement}%
  \BibitemOpen
  \bibfield  {author} {\bibinfo {author} {\bibfnamefont {J.~C.}\ \bibnamefont
  {Hoke~\textit{et al.,} {(Google Quantum AI and Collaborators)}}},\ }\bibfield
   {title} {\bibinfo {title} {Measurement-induced entanglement and
  teleportation on a noisy quantum processor},\ }\href
  {https://doi.org/10.1038/s41586-023-06505-7} {\bibfield  {journal} {\bibinfo
  {journal} {Nature}\ }\textbf {\bibinfo {volume} {622}},\ \bibinfo {pages}
  {481} (\bibinfo {year} {2023})}\BibitemShut {NoStop}%
\bibitem [{\citenamefont {Guo}\ \emph {et~al.}(2024)\citenamefont {Guo},
  \citenamefont {Wu}, \citenamefont {Ye}, \citenamefont {Zhang}, \citenamefont
  {Lian}, \citenamefont {Yao}, \citenamefont {Wang}, \citenamefont {Yan},
  \citenamefont {Yi}, \citenamefont {Xu}, \citenamefont {Li}, \citenamefont
  {Hou}, \citenamefont {Xu}, \citenamefont {Guo}, \citenamefont {Zhang},
  \citenamefont {Qi}, \citenamefont {Zhou}, \citenamefont {He},\ and\
  \citenamefont {Duan}}]{guo2024nature}%
  \BibitemOpen
  \bibfield  {author} {\bibinfo {author} {\bibfnamefont {S.-A.}\ \bibnamefont
  {Guo}}, \bibinfo {author} {\bibfnamefont {Y.-K.}\ \bibnamefont {Wu}},
  \bibinfo {author} {\bibfnamefont {J.}~\bibnamefont {Ye}}, \bibinfo {author}
  {\bibfnamefont {L.}~\bibnamefont {Zhang}}, \bibinfo {author} {\bibfnamefont
  {W.-Q.}\ \bibnamefont {Lian}}, \bibinfo {author} {\bibfnamefont
  {R.}~\bibnamefont {Yao}}, \bibinfo {author} {\bibfnamefont {Y.}~\bibnamefont
  {Wang}}, \bibinfo {author} {\bibfnamefont {R.-Y.}\ \bibnamefont {Yan}},
  \bibinfo {author} {\bibfnamefont {Y.-J.}\ \bibnamefont {Yi}}, \bibinfo
  {author} {\bibfnamefont {Y.-L.}\ \bibnamefont {Xu}}, \bibinfo {author}
  {\bibfnamefont {B.-W.}\ \bibnamefont {Li}}, \bibinfo {author} {\bibfnamefont
  {Y.-H.}\ \bibnamefont {Hou}}, \bibinfo {author} {\bibfnamefont {Y.-Z.}\
  \bibnamefont {Xu}}, \bibinfo {author} {\bibfnamefont {W.-X.}\ \bibnamefont
  {Guo}}, \bibinfo {author} {\bibfnamefont {C.}~\bibnamefont {Zhang}}, \bibinfo
  {author} {\bibfnamefont {B.-X.}\ \bibnamefont {Qi}}, \bibinfo {author}
  {\bibfnamefont {Z.-C.}\ \bibnamefont {Zhou}}, \bibinfo {author}
  {\bibfnamefont {L.}~\bibnamefont {He}},\ and\ \bibinfo {author}
  {\bibfnamefont {L.-M.}\ \bibnamefont {Duan}},\ }\bibfield  {title} {\bibinfo
  {title} {A site-resolved two-dimensional quantum simulator with hundreds of
  trapped ions},\ }\href {https://doi.org/10.1038/s41586-024-07459-0}
  {\bibfield  {journal} {\bibinfo  {journal} {Nature}\ }\textbf {\bibinfo
  {volume} {630}},\ \bibinfo {pages} {613} (\bibinfo {year}
  {2024})}\BibitemShut {NoStop}%
\bibitem [{Note1()}]{Note1}%
  \BibitemOpen
  \bibinfo {note} {The converse, however, is not generally true; certain
  entangled states may also yield zero or even negative values for the reduced
  fluctuations.}\BibitemShut {Stop}%
\bibitem [{\citenamefont {Fishman}\ \emph
  {et~al.}(2022{\natexlab{a}})\citenamefont {Fishman}, \citenamefont {White},\
  and\ \citenamefont {Stoudenmire}}]{itens1}%
  \BibitemOpen
  \bibfield  {author} {\bibinfo {author} {\bibfnamefont {M.}~\bibnamefont
  {Fishman}}, \bibinfo {author} {\bibfnamefont {S.~R.}\ \bibnamefont {White}},\
  and\ \bibinfo {author} {\bibfnamefont {E.~M.}\ \bibnamefont {Stoudenmire}},\
  }\bibfield  {title} {\bibinfo {title} {{The ITensor Software Library for
  Tensor Network Calculations}},\ }\href
  {https://doi.org/10.21468/SciPostPhysCodeb.4} {\bibfield  {journal} {\bibinfo
   {journal} {SciPost Phys. Codebases}\ ,\ \bibinfo {pages} {4}} (\bibinfo
  {year} {2022}{\natexlab{a}})}\BibitemShut {NoStop}%
\bibitem [{\citenamefont {Fishman}\ \emph
  {et~al.}(2022{\natexlab{b}})\citenamefont {Fishman}, \citenamefont {White},\
  and\ \citenamefont {Stoudenmire}}]{itens2}%
  \BibitemOpen
  \bibfield  {author} {\bibinfo {author} {\bibfnamefont {M.}~\bibnamefont
  {Fishman}}, \bibinfo {author} {\bibfnamefont {S.~R.}\ \bibnamefont {White}},\
  and\ \bibinfo {author} {\bibfnamefont {E.~M.}\ \bibnamefont {Stoudenmire}},\
  }\bibfield  {title} {\bibinfo {title} {{Codebase release 0.3 for ITensor}},\
  }\href {https://doi.org/10.21468/SciPostPhysCodeb.4-r0.3} {\bibfield
  {journal} {\bibinfo  {journal} {SciPost Phys. Codebases}\ ,\ \bibinfo {pages}
  {4}} (\bibinfo {year} {2022}{\natexlab{b}})}\BibitemShut {NoStop}%
\bibitem [{\citenamefont {Gross}\ \emph {et~al.}(2010)\citenamefont {Gross},
  \citenamefont {Liu}, \citenamefont {Flammia}, \citenamefont {Becker},\ and\
  \citenamefont {Eisert}}]{Eisert2010PRL}%
  \BibitemOpen
  \bibfield  {author} {\bibinfo {author} {\bibfnamefont {D.}~\bibnamefont
  {Gross}}, \bibinfo {author} {\bibfnamefont {Y.-K.}\ \bibnamefont {Liu}},
  \bibinfo {author} {\bibfnamefont {S.~T.}\ \bibnamefont {Flammia}}, \bibinfo
  {author} {\bibfnamefont {S.}~\bibnamefont {Becker}},\ and\ \bibinfo {author}
  {\bibfnamefont {J.}~\bibnamefont {Eisert}},\ }\bibfield  {title} {\bibinfo
  {title} {Quantum state tomography via compressed sensing},\ }\href
  {https://doi.org/10.1103/PhysRevLett.105.150401} {\bibfield  {journal}
  {\bibinfo  {journal} {Phys. Rev. Lett.}\ }\textbf {\bibinfo {volume} {105}},\
  \bibinfo {pages} {150401} (\bibinfo {year} {2010})}\BibitemShut {NoStop}%
\bibitem [{\citenamefont {Huang}\ \emph {et~al.}(2020)\citenamefont {Huang},
  \citenamefont {Kueng},\ and\ \citenamefont {Preskill}}]{huang2020predicting}%
  \BibitemOpen
  \bibfield  {author} {\bibinfo {author} {\bibfnamefont {H.-Y.}\ \bibnamefont
  {Huang}}, \bibinfo {author} {\bibfnamefont {R.}~\bibnamefont {Kueng}},\ and\
  \bibinfo {author} {\bibfnamefont {J.}~\bibnamefont {Preskill}},\ }\bibfield
  {title} {\bibinfo {title} {Predicting many properties of a quantum system
  from very few measurements},\ }\href
  {https://doi.org/10.1038/s41567-020-0932-7} {\bibfield  {journal} {\bibinfo
  {journal} {Nature Physics}\ }\textbf {\bibinfo {volume} {16}},\ \bibinfo
  {pages} {1050} (\bibinfo {year} {2020})}\BibitemShut {NoStop}%
\bibitem [{\citenamefont {G\l{}odzik}\ \emph {et~al.}(2025)\citenamefont
  {G\l{}odzik}, \citenamefont {Ghorbanzadeh~Moghaddam}, \citenamefont
  {P\"oyh\"onen},\ and\ \citenamefont {Ojanen}}]{zenodo}%
  \BibitemOpen
  \bibfield  {author} {\bibinfo {author} {\bibfnamefont {S.}~\bibnamefont
  {G\l{}odzik}}, \bibinfo {author} {\bibfnamefont {A.}~\bibnamefont
  {Ghorbanzadeh~Moghaddam}}, \bibinfo {author} {\bibfnamefont {K.}~\bibnamefont
  {P\"oyh\"onen}},\ and\ \bibinfo {author} {\bibfnamefont {T.}~\bibnamefont
  {Ojanen}},\ }\bibfield  {title} {\bibinfo {title} {Entanglement entropy
  scaling laws from fluctuations of non-conserved quantities},\ }\href
  {https://doi.org/10.5281/zenodo.16276293} {10.5281/zenodo.16276293} (\bibinfo
  {year} {2025})\BibitemShut {NoStop}%
\end{thebibliography}%
\end{document}